\documentclass{aa}

\usepackage{graphicx}

\begin{document}

\title{A near--infrared survey for Galactic Wolf-Rayet stars}

\author{Nicole L. Homeier \inst{1}\thanks{Visiting Astronomer, Cerro Tololo Inter-American Observatory, National Optical Astronomy Observatories, which are operated by the Association of Universities for Research in Astronomy, Inc., under cooperative agreement with the National Science Foundation.} \and Robert D. Blum \inst{2}* \and Peter S. Conti \inst{3} \and Augusto Damineli \inst{4}{\scriptsize*}}

\offprints{N. Homeier, \email{nhomeier@eso.org}}

\institute{European Southern Observatory, Karl Schwarzschild Str. 2, Garching bei Muenchen \and Cerro Tololo Interamerican Observatory, Casilla 603, La Serena, Chile \and JILA and APS Department, University of Colorado, Boulder, CO 8030 \and IAG, S\~{a}o Paulo, Brazil}

\date{Received  Sept. 12, 2002/ Accepted Oct. 23, 2002}

\titlerunning{NIR Survey for Galactic WRs}
\authorrunning{Homeier et al.}

\abstract{Initial results, techniques, and rationale for a
near--infrared survey of evolved emission--line stars toward the
Galactic Center are presented.  We use images taken through
narrow-band emission--line and continuum filters to select candidates
for spectroscopic follow--up. The filters are optimized for the
detection of Wolf-Rayet stars and other objects which exhibit
emission--lines in the 2 $\mu$m region.  Approximately three square
degrees along the Galactic plane have been analyzed in seven
narrow--filters (four emission--lines and three continuum). Four new
Wolf--Rayet stars have been found which are the subject of a following
paper.}

\maketitle

\keywords{Stars: Wolf-Rayet, Galaxy: stellar content, Galaxy: center, Infrared: Stars}

\section{Introduction}

Optical surveys within our Galaxy are severely hampered by dust obscuration;
therefore complete samples must be obtained with 
longer wavelength studies. Here we describe a survey procedure
for evolved massive stars in the Galactic
plane at K-band wavelengths, where the extinction is 10 magnitudes
lower than for traditional V-band surveys. 
Our scientific driver is the discovery of young stellar populations
in our Galaxy through the detection of evolved massive stars. These
stars have strong emission lines, which makes them relatively easy to 
detect using narrow-band filters.

Massive stars drive the evolution of galaxies through powerful stellar
winds, large mass ejections, and explosive deaths. These
mechanisms are the dominant source of energy input into the ISM;
 thus the structure and composition of the ISM in most galaxies is
largely determined by the massive star population
(\cite{LRD92}, \cite{OC97}, \cite{Hetal98}, \cite{M99}, 
\cite{Oetal01}, \cite{Hetal01}). 
Massive stars are also
essential contributors to the chemical evolution of their host 
galaxy, ejecting material enriched in helium, carbon, and nitrogen during
their lives, and depositing elements
heavier than nitrogen in their final eruption as SNe.

As a massive star evolves, its spectrum becomes dominated by emission
lines, arising either in a dense stellar wind, or in circumstellar
material produced by mass loss. The presence and strength of
individual lines are clues to the star's evolutionary state and
atmospheric structure. Among evolved massive stars with such spectra,
emission lines are most pronounced in Wolf-Rayet
(WR) stars and in the Luminous Blue Variable (LBV, or S Dor variable) 
stage, where they shed large (1 to 10 M$_{\odot}$) 
amounts of chemically enriched matter in a relatively small amount of
time ($\sim10^{4}$yrs) (\cite{Petal97}, \cite{Setal98}, \cite{Letal99}).

WR stars have lifetimes $<$~10~Myr, and thus are excellent tracers of recent
star formation, and so also Galactic structure. They are also important 
in our quest to understand how star formation proceeds. For example, most
of the previously known WRs are relatively isolated or in OB associations, 
but recent searches in the IR have found a plethora of these objects in
compact clusters near the Galactic center (\cite{BDS95}, \cite{Ketal95}, 
\cite{Netal95}, \cite{Fetal99a}, \cite{Betal01}). 

\section{Life Cycle of a Massive Star}

In the 'Conti scenario', the evolution of a massive star is determined
by the amount of mass loss it undergoes (\cite{C76}, \cite{MC94}). This
mass loss is driven by the combination of the star's intrinsic luminosity 
and opacity due to metal lines. The more luminous a star is, the greater the 
outward force, and the more metals present in the wind, the greater the 
opacity and hence the ability to drive a mass-losing stellar wind. 
Our picture of massive stellar evolution is rapidly changing, as new 
models including rotation are developed (\cite{MM00a},
\cite{MM00b}). It now appears that rotation is second only to mass loss
rate in its effect on massive stellar evolution, and it may even be 
{\it more} important at low metallicities (\cite{MM01}). 

The current qualitative overview of massive stellar 
evolution is as follows. For a star with an initial mass of 
$\ge 30$ M$_{\odot}$, mass loss and mixing on the main sequence deplete 
the hydrogen-rich envelope and
reveal the equilibrium products of CNO-cycle hydrogen
burning, creating a WN star. Additional mass loss and mixing reveals
the products of He-burning, and the star becomes a WC star. The 
star will end its life as a Type Ib or Ic supernova. This evolution may
or may not be punctuated by eruptions and episodes of large mass loss where 
the star is identified as an LBV.

\section{The need for NIR observations}

The ideal place to study massive star formation and 
evolution is our own Galaxy, where we can
resolve objects on small linear scales and have some hope of 
complete samples. It would be of great interest to find 
{\it all} the WR stars in our galaxy to test stellar evolution theories 
at high metallicity, and to learn about environments of massive star 
formation. Another important benefit is the study of 
Galactic structure as traced by young star-forming regions.

Surveys for emission line stars in the Milky Way using the narrow band
technique have been done before, but in the optical, where
extinction by intervening dust is highly problematic (\cite{Setal99}). 
However, these
surveys firmly establish WR stars as key tracers of star formation,
both in the Milky Way and other galaxies (\cite{C91}). By
moving to the NIR, we can identify these objects
in regions where heavy reddening renders them undetectable in the optical.

With a simple model for the distribution of Galactic WR stars, 
Shara et al. (1999),
demonstrate the difficulty in conducting WR searches in the
optical. Because of high extinction in the plane,
the vast majority ($\sim 90$\%) of the Milky Way's
massive stellar population is hidden from view by obscuring dust. 
They also showed how this could be overcome by moving to the
infrared, indicating that the number of WR stars as a
function of magnitude peaks at K$=13-14$. It is clear that 
NIR surveys within the plane
of our Galaxy are essential to significantly enlarge the known population
of WR stars.

\section{Survey Description}

\begin{table}
\caption[]{Filter Description}
\begin{center}
\begin{tabular}{cc}\hline\hline
Central $\lambda$ ($\mu$m) & FWHM ($\mu$m) \\\hline
2.032 & 0.010  \\
2.062 & 0.010  \\
2.077 & 0.015 \\
2.142 & 0.020  \\
2.161 & 0.022 \\
2.191 & 0.013 \\
2.248 & 0.024 \\\hline
\end{tabular}
\end{center}
\end{table}

\begin{table}
\caption[]{Prominent Lines in Wolf-Rayet K-band Spectra}
\begin{flushleft}
\begin{tabular}{lll}\hline\hline
Central $\lambda$ ($\mu$m) & Transition & Type\\\hline
2.0587 & He I $2s-2p$ & WN, WC \\
2.0705/2.0796/2.0842  & C IV  $3p-3d$ & WC \\
2.1126/2.1137 & He I $4s-3p$ & WN, WC \\
2.1038/2.1152/2.1155/2.1156  & C III/N III $8-7$ & WC \\
2.1632 & He I $7-4$ & WN, WC \\
2.1652 & He II $14-8$ & WN, WC \\
2.166 & Br $\gamma$, H I $7-4$ & WN \\
2.189 & He II $10-7$ & WN, WC \\
2.2779 & C IV $15-12$ & WC \\
2.3178 & C IV $17-13$ & WC \\
2.3470 & He II $13-8$ & WN \\\hline
\end{tabular}
\begin{list}{}{}
\item Data is reproduced from Table 2 of Figer et al. (1997), see references therein. Wavelengths listed are vacuum wavelengths.
\end{list}
\end{flushleft}
\end{table}

\begin{figure*}
 \parbox{80.mm}{
 \centering
 \includegraphics[width=8.5cm]{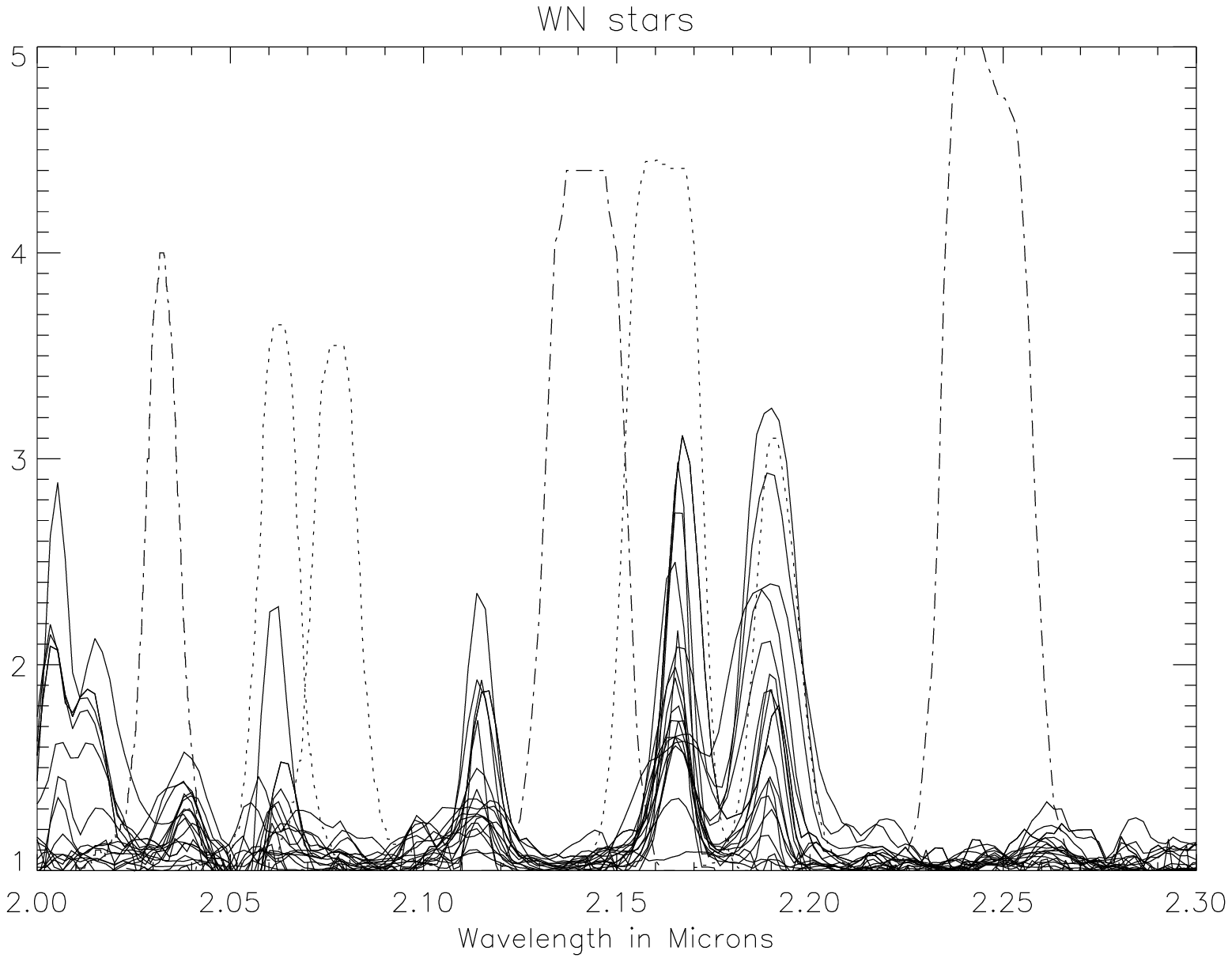}
}
\hspace{10.0mm}
\parbox{80.mm}{
 \includegraphics[width=8.5cm]{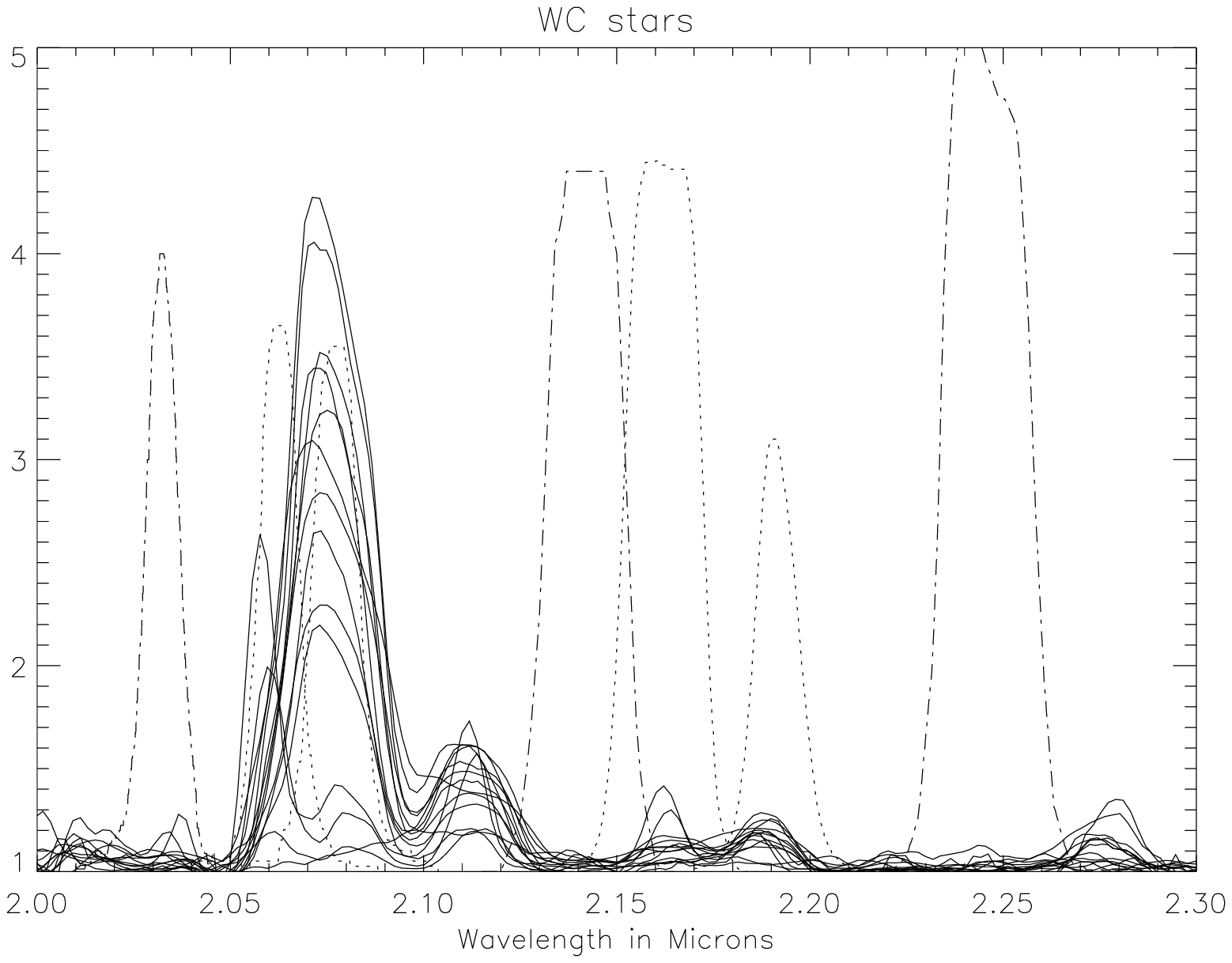}
}
\caption{A representative sample of WR stars with filter transmission
curves overplotted, WN stars {\it left} panel and WC stars {\it right}
panel. Dotted curves represent the line filters at 2.06, 2.08, 2.166,
and 2.19 $\mu$m. Dot--dashed lines represent the continuum filters at
2.03, 2.14, and 2.248 $\mu$m. The He~II filter transmission curve is
difficult to see in the left panel due to the strong He II lines in the
WN stars. WR spectra in both panels are kindly provided by P. Eenens.}
\label{fo}
\end{figure*}

In Table 1 we present central wavelengths and FWHMs for
our chosen set of K-band filters (\cite{BD99a}, \cite{Hetal02}).
Four filters are centered on 
the characteristic stellar wind emission lines of He I 2.06 $\mu$m, 
C IV 2.08 $\mu$m, H I Br$\gamma$ 2.166 $\mu$m,
and He II 2.189 $\mu$m, and the additional three continuum filters
are at 2.03 $\mu$m, 2.14 $\mu$m, and 2.248 $\mu$m. Thus each line filter
measurement has a continuum point to the red and blue. 
It is extremely important to have continuum points to the
red and the blue of each line filter because of the variation in continuum
slope caused by dust extinction.

In Table 2 we list the most prominent emission lines in WR
spectra, and the filter response curves are overplotted in 
Figures \ref{fo}a and 
\ref{fo}b for unpublished $K-$band spectra of WN and WC stars kindly 
provided by P. Eenens. 
In both, the three continuum filters are overplotted as 
dot--dash lines, and the four line filters as dotted lines. This illustrates
the sensitivity of the 2.17 and 2.19 filters to WN stars, whereas
the 2.06 and 2.08 filters are most sensitive to WC stars. 

\subsection{Observations and Reductions}

\begin{figure*}
 \centering
 \includegraphics[width=17cm]{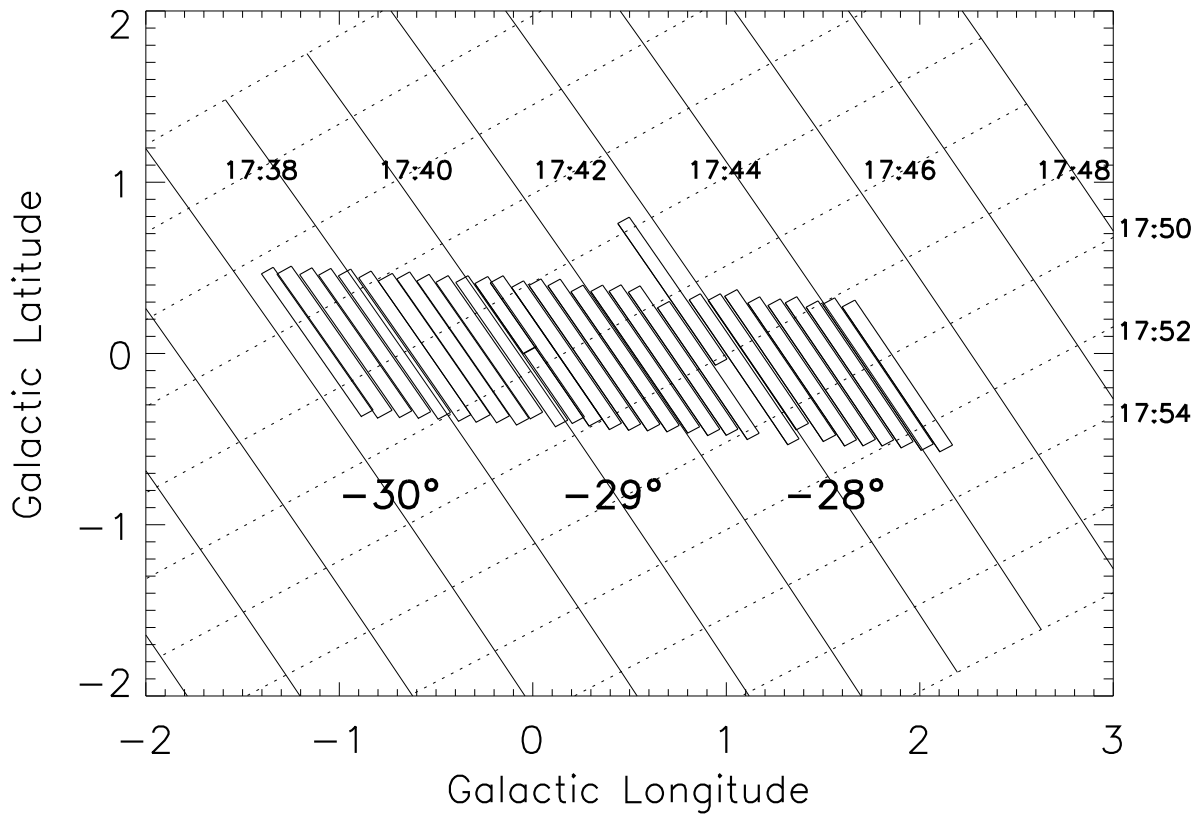}
 \caption{Map of
 the observations from 1996, 1997, and 1998.  The x and y axes are in
 Galactic coordinates. Lines of constant right ascension are overplotted in
 intervals of 2 minutes as dotted lines, and lines of constant
 declination are overplotted every half degree as solid lines.}
 \label{obsgrid}
\end{figure*}

\begin{figure*}
 \centering
 \caption{The same data as in Figure \ref{obsgrid} overplotted on the 90cm 
radio image of the Galactic Center region presented in LaRosa et al. 2000.
Major features are labeled, and the small rectangle indicates the approximate
position of Figure \ref{data}.}
 \label{90cmb}
\end{figure*}

\begin{figure*}
 \centering
 \includegraphics[width=17cm]{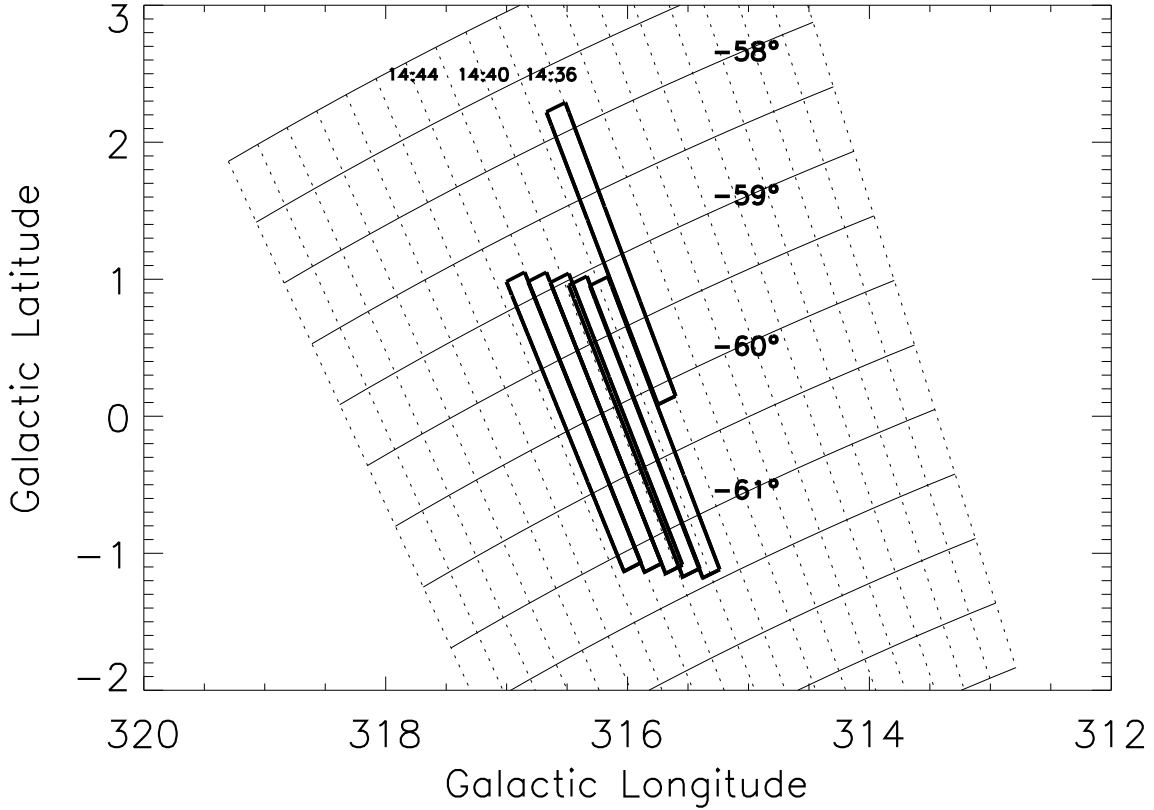}
 \caption{Same as Figure~\ref{obsgrid}, but for the 2000 data.}
 \label{obsgrid2000}
\end{figure*}

\begin{figure*}
 \centering
 \caption{This is a sample of our data from 1996. The image shown here was
taken with the 2.248 $\mu$m filter. The center coordinates
are 17:46:35.7, -28:42:35 J2000. The y dimension spans 
$5 \arcmin$ and the
x dimension spans $\sim 19 \arcmin$. South is up and east is to the
left. Large scale variations in stellar density are apparent and caused by 
intervening dark clouds.}
 \label{data}
\end{figure*}

In 1996 the survey was begun at the 1.5m at Cerro Tololo Interamerican
Observatory, and continued in 1997, 1998, and 2000. For 1996, 97, and
98, we used the same telescope and instrument configuration: the Cerro
Tololo Infrared Imager (CIRIM) with the f/8 mode, giving us a $5
\arcmin \times 5 \arcmin$ field of view and $1.16 \arcsec$ per
pixel. In 2000, we used the Ohio State Infrared Imaging Spectrometer
in the f/8 mode for a $10 \arcmin \times 10 \arcmin$ field of view and
$1.16 \arcsec$ per pixel.  Our images are taken in 'strips' composed
of $35-45$ images.  We offset $1/3$ of a chip in RA (1996--1997 data)
or Dec (2000 data), keeping the other coordinate constant. Thus each
spot on the image strip is a composite of three exposures. Images have
been obtained over $\sim 3$~degrees of longitude between $\pm 0.5$
near the Galactic Center. This is shown in Figure \ref{obsgrid}, and
overplotted on a 90cm image of the Galactic Center region in Figure
\ref{90cmb}. In 2000, we moved outward along the plane to $l = 316$, at
the edge of Centaurus looking towards the Scutum--Crux spiral arm. In
this region we obtained $\sim 1$ degree of Galactic longitude between $\pm
1.0$ Galactic Latitude. This is shown in Figure \ref{obsgrid2000}. A sample
of our data is shown in Figure \ref{data}, with a small rectangle
indicating the approximate position on the 90cm radio image in Figure
\ref{90cmb}. 

Data reduction is performed with the CIRRED package of routines in 
IRAF\footnote{IRAF is distributed by the National Optical Astronomy 
Observatories},
written specifically for CIRIM and OSIRIS\footnote{OSIRIS is a collaborative 
project between the Ohio State University and CTIO. OSIRIS was developed
through NSF grants AST 90--16112 and AST 92--18449} reductions by 
R. D. Blum. The 
reduction steps are as follows. First, we trim the flat-field images if 
needed. For each filter there is a set of images taken with the lamps
on and off. We take the median of each set of 'on' and 'off' frames to make
a single 'on' and 'off' image. We subtract
the 'off' frame from the 'on' frame to make a flat image, which is then 
normalized by the mean. A bad pixel mask is made by comparing the dome
flats with the lights on and the lights off and using the histogram of 
pixel intensities to distinguish good and bad
pixels. This bad pixel mask is then used to correct the flat field
images.

Next, the survey images are trimmed if needed, linearity corrected
with IRLINCOR, divided by exposure time, fixed with the bad pixel
mask, and divided by the flat field image. Then the entire stack of
images within a strip is used to make a sky image for each filter. The
images are median combined using 'minmax' rejection with approximately half 
the images thrown out to reject contributions
from stars. The resulting sky image is subtracted from each of the
individual frames, and a constant is added back to maintain an
appropriate sky level in the reduced image. These images are then
assembled into a strip approximately one degree long by
cross--correlation of the overlapping regions on each frame. Finally,
we derive astrometric solutions for each of our strips by comparing
the 248 filter images to the 2MASS catalog images from IPAC and the
IRAF task CCMAP.

\subsection{Photometry}

Each frame is analyzed with DoPhot (\cite{SMS93}). DoPhot
identifies, classifies, and performs photometry on objects in an
image. It makes successive passes over the image, subtracting the
objects and searching in the next pass for fainter objects. The model
parameters are found iteratively in the image itself, with reasonable first
guesses supplied by the user in parameter files.

\subsection{Apparent Magnitude Comparison}

Our program has no strict requirement for calibrated photometry since
emission--line stars are found through continuum independent line
indices. Furthermore, much of the data was taken in non--photometric
conditions.  However, it is of interest to have an
order--of--magnitude calibration in order to help assess how deep the
images typically go.  We have done a comparison between a typical
image section of one of our image strips and the 2MASS survey images.
We compared our instrumental magnitude for the 2.06 $\mu$m filter,
which is the least sensitive of all our filters due to the combination of
transmission efficiency and filter width. 2MASS magnitudes were
obtained from the Infrared Science Archive (IRSA) through the GATOR
query page.  Our comparison shows that an instrumental magnitude of
$-3$ corresponds roughly to an apparent $K$ magnitude of 12 for this
strip. 

\subsection{Completeness}

\begin{figure*}
 \centering
 \includegraphics[width=17cm]{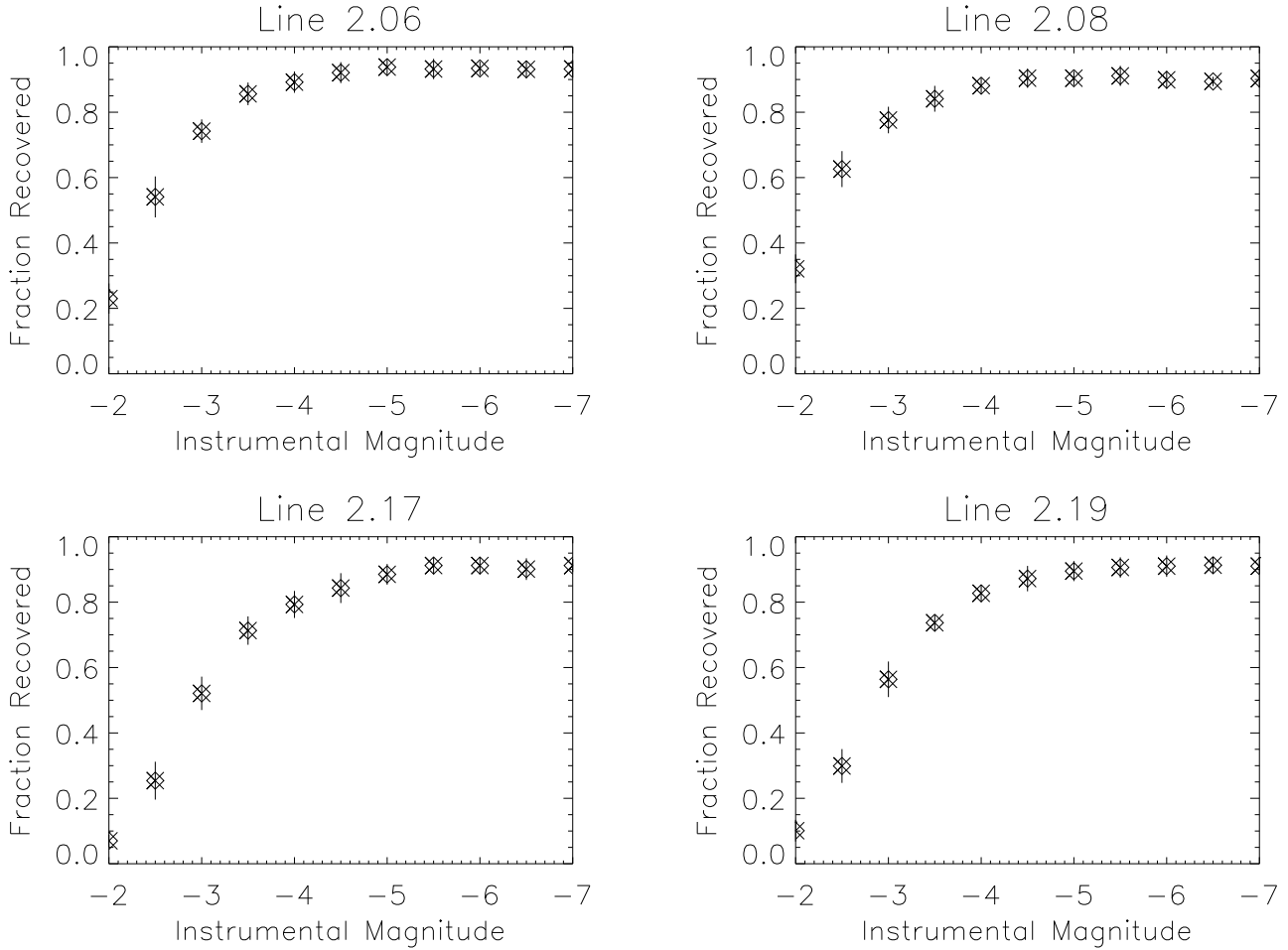}
 \caption{The
 fraction of stars recovered in artificial star experiments for the
 four line images and its dependence on the continuum magnitude of the
 star. See text.}
 \label{frac}
\end{figure*}

\begin{figure*}
 \centering
 \includegraphics[width=17cm]{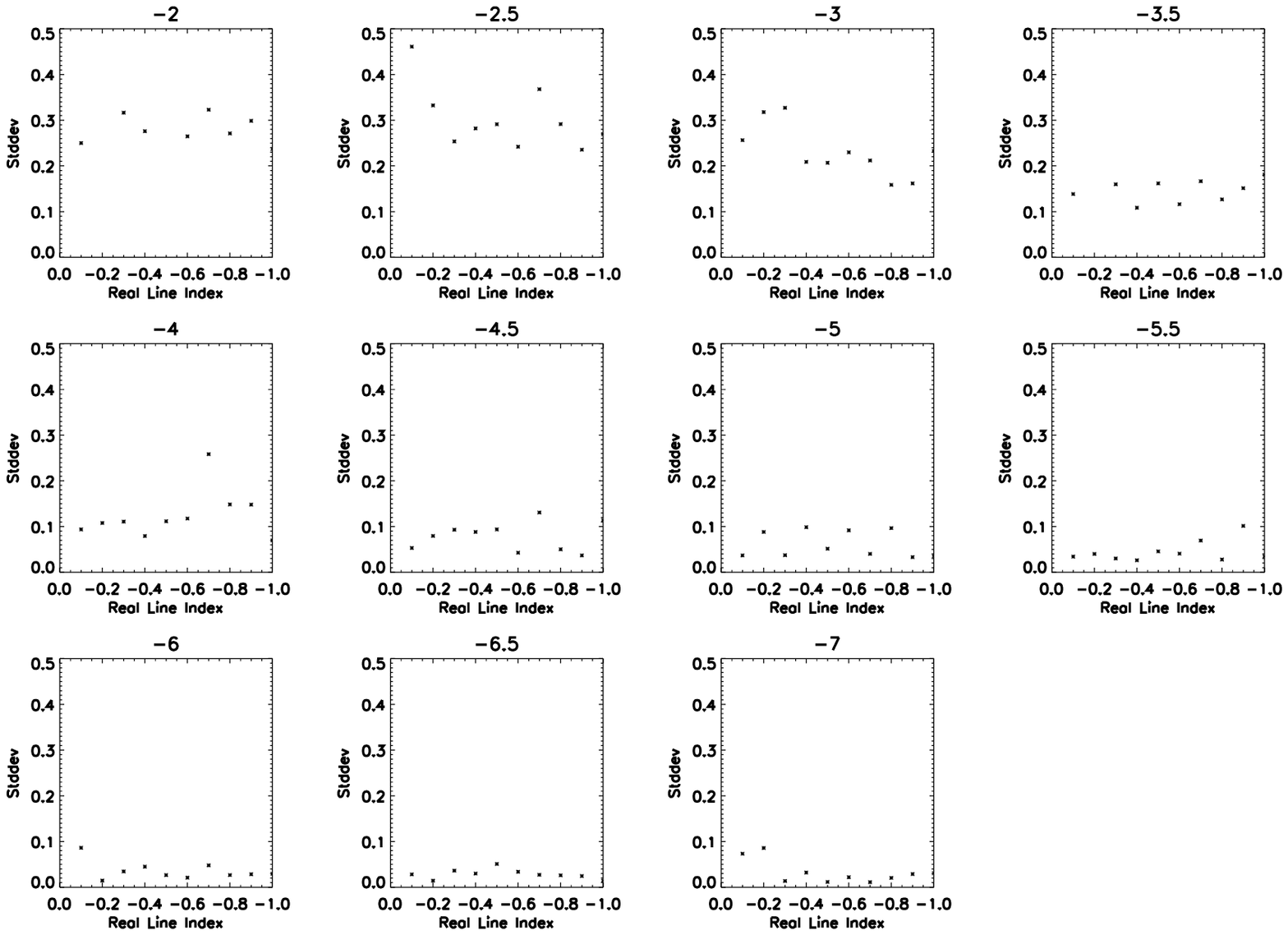}
 \caption{Standard
 deviation of the difference between the input color index and the
 measured line index for the 2.06 line artificial star experiment. The
 continuum magnitude is shown on the top of each small panel. The error
 in line index does not depend on line strength, but does depend on
 continuum magnitude.}
 \label{rms}
\end{figure*}

To test our sensitivity to emission line stars, we created grids of
fake stars with continuum magnitudes ranging from -2.0 to -7.0 in 
instrumental continuum magnitude, and
emission line magnitudes ranging from 0.1 to 1.0 magnitudes 
brighter than the continuum. That is, we created grids where 100 stars have
continuum (2.03, 2.14, and 2.248 $\mu$m) magnitudes of -2.0, and 
line (2.062, 2.077, 2.166, 2.191 $\mu$m) magnitudes of -2.1, 100 with the 
same continuum magnitudes, and line magnitudes of -2.2, and so on.
At each step in continuum magnitude, we added 1000 stars with varying
line magnitude. 

The psf for each filter image was constructed
using several bright and relatively well--isolated stars in the image 
and the IRAF tasks
'pstselect' and 'psf'. Images were then created using
'mkobjects'. These images were analyzed in exactly the same way as a real
image. In Figure \ref{frac} we plot
the fraction recovered at increasing brightness of the continuum magnitude. 
There is a dependence on overall brightness, but no significant dependence 
on emission line strength. This is similar to what one would expect
for a broad--band color.

The survey is limited by crowding (especially in the GC region) and
the relatively coarse angular resolution element provided by the 1.5m
telescope. Given the order--of--magnitude comparison made above with
the comparison to 2MASS, the 2.06 $\mu$m filter the data are
$\ge 75$~\% complete at $K =$ 12 mag. However, the photometric accuracy
decreases at faint magnitudes. This is shown in Figure \ref{rms} for
the same stars that are plotted in Figure \ref{frac}. The line shown
here is the $2.062 \mu$m line, but all lines (2.077, 2.166, and 2.191
$\mu$m) show very similar behavior.

\subsection{Candidate Selection}

In this section we detail our candidate selection procedure. As a first step,
the object coordinates must be transformed to a single system. We
have arbitrarily chosen to use the longest wavelength filter image as 
our template. 
Thus, we scale and offset the coordinates for objects in the other lists
to the 2.248 $\mu$m object list. This is accomplished with the 'transform' 
package which accompanies DoPhot. Transform uses a triangular search routine 
to match stars in different images by comparing the relative scales and
orientations of stars in groups of three. Hereafter, we refer to the task 
as 'offset'.

We look at each emission--line filter separately, meaning we consider sets
of three filters each time, the line and its two continuum filters 
(2.03 and 2.14 for 2.06 and 2.08; 2.14 and 2.248 for 2.17 and 2.19).
The raw output of DoPhot contains 10,000 to 30,000 objects per image,
and this output must first be gleaned for stars with good statistics.
The first cut is on object type. In the DoPhot output each object is 
assigned a number indicating the whether it is a star or a galaxy, and
how well it thinks it can determine the magnitude. There are three object
types we are interested in: 1's are single stars with good statistics, 7's
are single stars with decent statistics, and 3's are fit as 
members of a blend of 2 
single stars. During the transformation to 2.248 $\mu$m coordinates,
'offset' keeps track of the object types in both lists, and records it 
in the output file. We keep only combinations
of single stars (such as 11's, 17's, 71's, or 77's) and doubles (33's). We 
throw out stars which have been classified as a member of a double in
one image and a single star in another, as the photometry in this case
is very unreliable. 

After cleaning the lists for object type, we perform a cut on photometric 
error. The error is calculated simply as the errors of the individual
magnitudes added in quadrature.
The error cut is a free parameter in our selection routine, but we usually 
select it to be between 0.1 and 0.3 magnitudes.

We now have three lists of objects, cleaned on the basis of object type and
error. The coordinates in these three lists are now matched, and a 
'line index' is computed as the magnitude
difference between the measured value and the expected value derived 
from a linear interpolation using the two continuum filters magnitudes. 

Essentially all stars in the images should have a constant line index
(which is not zero due to filter transmission differences and sky
transmission variations), that we can subtract to set the mean line
index to zero. When the nights were not photometric,
systematic line index variations as a function of position
(time) along the strip occur. These are relatively straightforward to 
account for since they produce large scale changes in the line indices with
position on the image. In our analysis we subtract these variations
by calculating a ``mean neighbor index'' for each star, using a bin
centered on the star, with a size between $50-100$ pixels. 
Smaller bins produce output with less scatter, but the bin must be large 
enough to contain stars to compute a significant ``neighbor index''.
This mean value is then subtracted from the individual objects.

\begin{figure}[!bh]
 \centering
 \includegraphics[width=8.5cm]{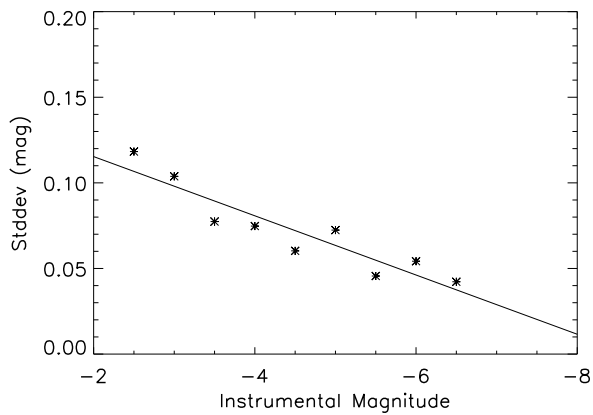}
 \caption{The scatter as a function of $2.03 \mu$m magnitude is fit
with a linear function to aid in candidate selection.}
 \label{stddev}
\end{figure}

We also consider the scatter in line index
as a function of the short wavelength continuum magnitude. An example 
is shown in Figure \ref{stddev}, where the standard deviation in
line index is fit as a linear function of short wavelength continuum 
magnitude. Our final selection is done by 
considering the line index, the error on this index, and the scatter in
line index at this magnitude. We employ a criterion similar to the 'S'
parameter employed by Damineli et al. (1997).

\begin{figure*}
 \parbox{80.mm}{
 \centering
 \includegraphics[width=8.5cm]{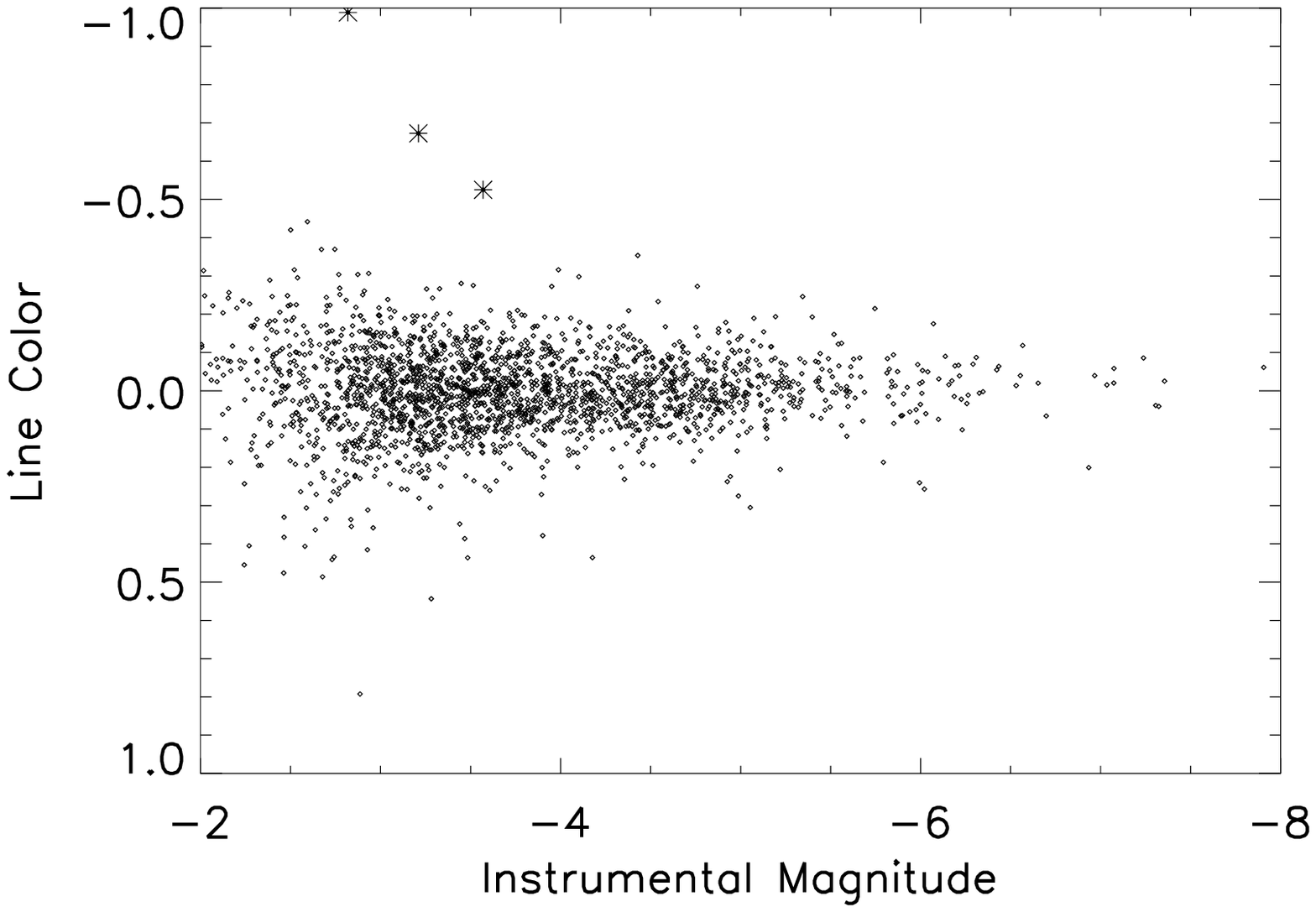}
}
\hspace{10.0mm}
\parbox{80.mm}{
 \includegraphics[width=8.5cm]{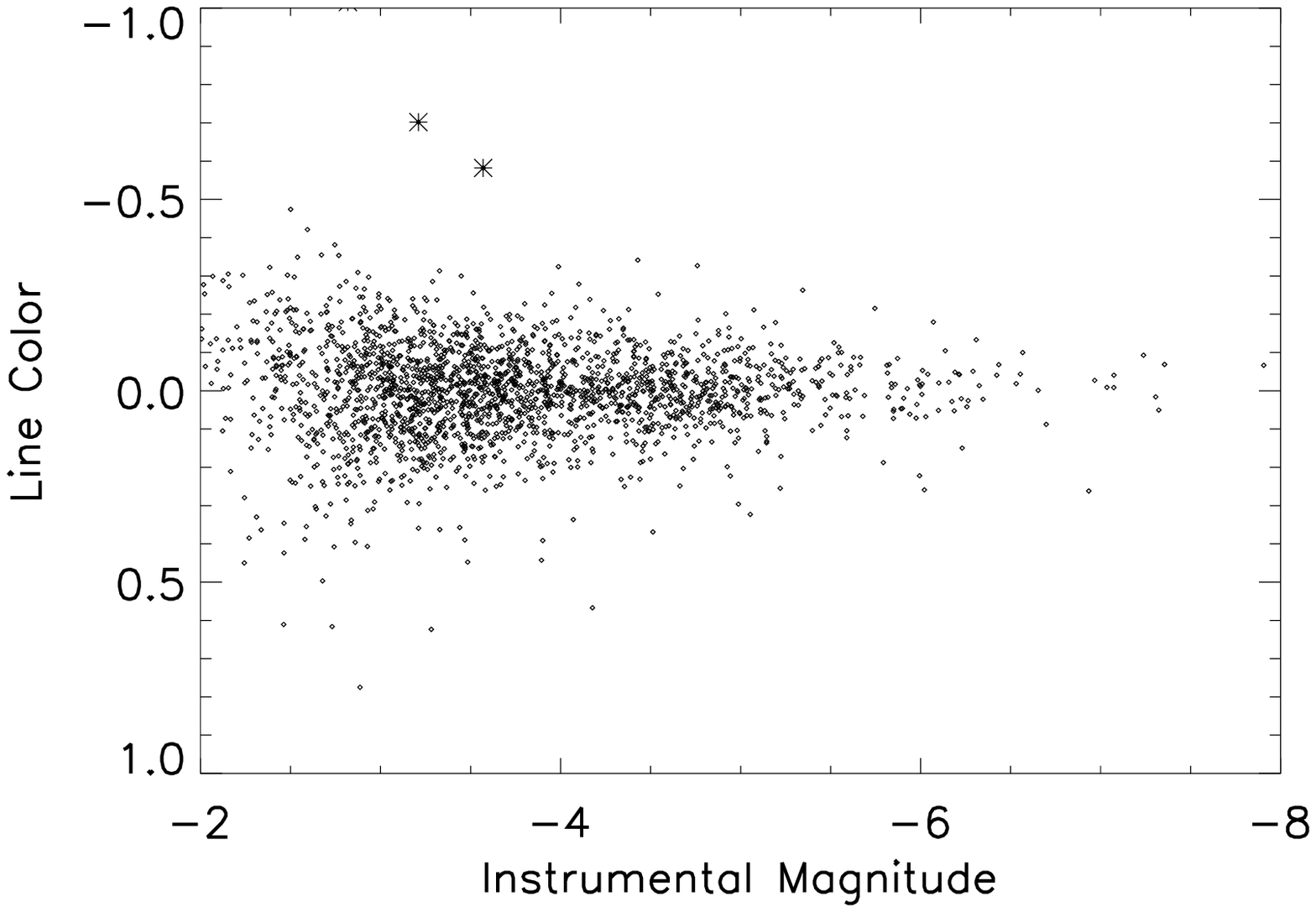}
}
\caption{2.06 Filter: An example of our photometric data and candidate 
selection. The
stars marked with asterisks are emission line candidates, and one is a 
confirmed WR star (cindex $\sim -0.6$, mag $\sim -3.7$; \cite{Hetal02}).
In the right panel is the same but without correcting for positional 
line index variations; see text.}
\label{magindex206}
\end{figure*}
\begin{figure*}
 \parbox{80.mm}{
 \centering
 \includegraphics[width=8.5cm]{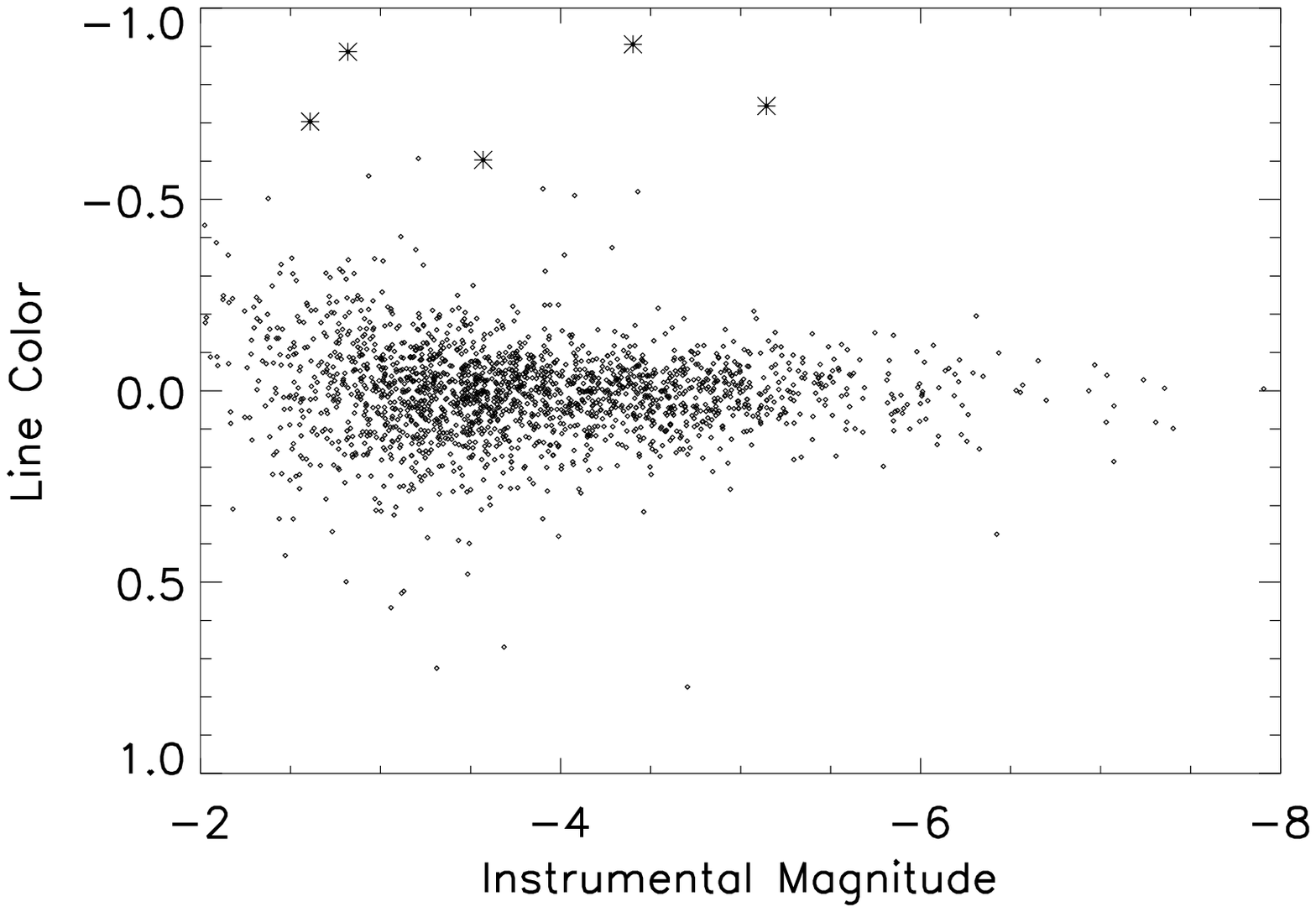}
}
\hspace{10.0mm}
\parbox{80.mm}{
 \includegraphics[width=8.5cm]{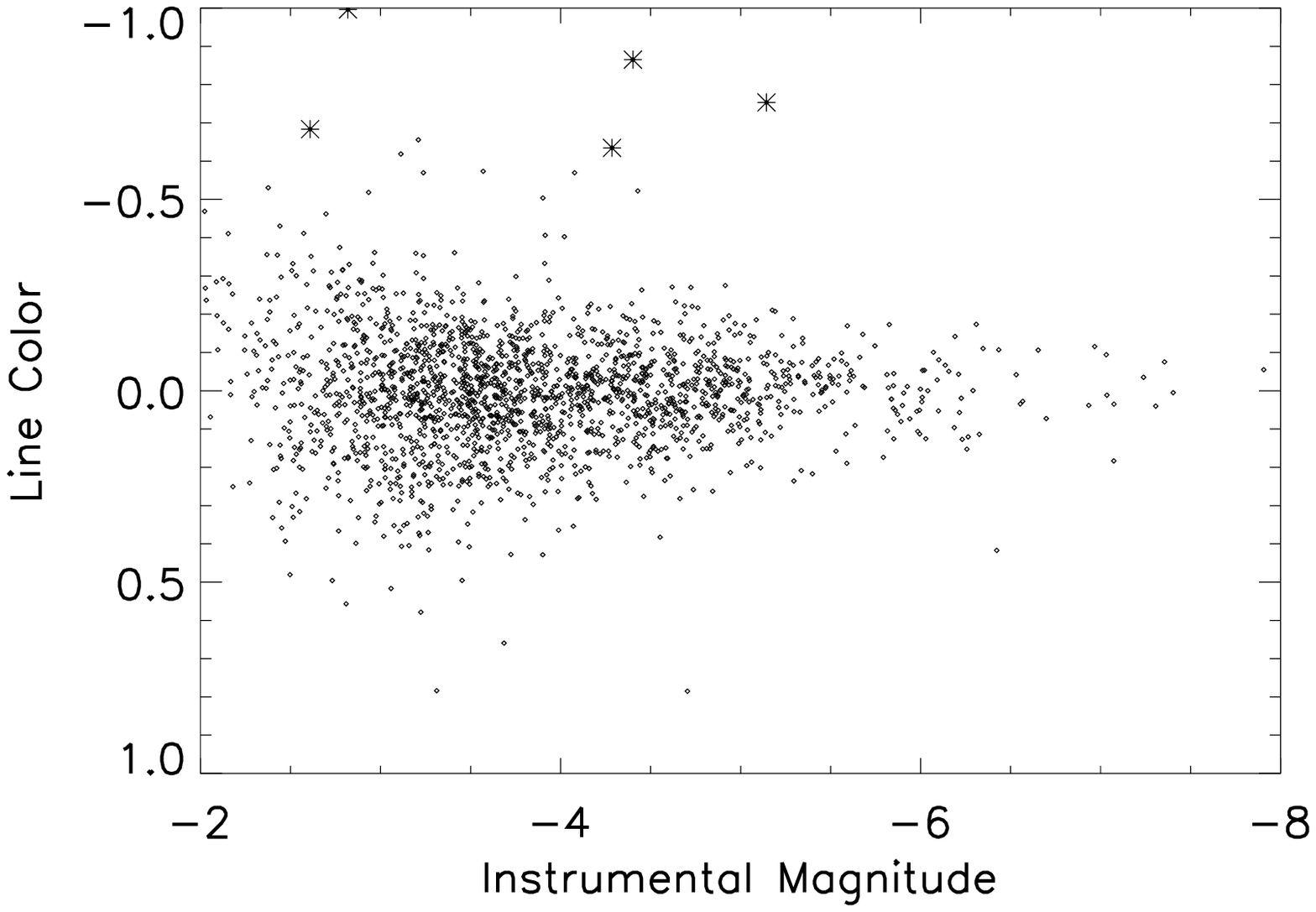}
}
\caption{Same as Figure~\ref{magindex206}, but for the $2.08 \mu$m filter.
The confirmed WR star is located at index $\sim -0.6$, mag $\sim
-3.7$.}
\label{magindex208}
\end{figure*}

An example of our photometry and candidate selection is shown in
Figures \ref{magindex206} and \ref{magindex208}. In these figures,
candidates were chosen as those with $k \geq 0.5$, and these
candidate emission line objects are overplotted as asterisks. One is a
confirmed WR star of late WC subtype (line index $\sim -0.6$,
inst. mag. $\sim -3.7$) (\cite{Hetal02}). In right hand panel of each
figure is the the same data plotted, but without correcting for
systematic positional variations in line index as described above. One
can see that the apparent scatter is larger for the objects in the
right-hand panels, and in the case of the 2.08 filter, it affects which
objects are chosen as emission-line candidates. In this particular
case, a bona-fide WR star would be missed at this cut on $k$.

\subsection{Image Subtraction}

We also use image subtraction as a way to select high priority candidates.
This step is only performed for those objects that are
selected as photometric candidates.
All procedures discussed here are accomplished with standard IRAF routines.
An image section is selected within $+/-$ 125 pixels of a photometric
candidate's x position, and a y pixel range from 1 to 250 pixels. First
the images are cross--correlated and shifted to a common position. 
The two continuum filters
are averaged and this result is normalized by the mean. The line
filter is also normalized by the mean, and then divided by the averaged,
normalized continuum image. The result is displayed on the screen and 
examined visually. We have found that this provides a useful and necessary 
check on the photometric selection. Effects such as bad pixels and
residual images are easily seen with this method.
We will discuss candidate selection in further detail in the upcoming paper
on the spectroscopic follow--up (Homeier et al. 2003, in prep).

\section{Estimating the Number of Candidates Expected}

\subsection{A Smooth Model}

Following the simple model of Shara et al. (1999), we can estimate the number 
of candidates we expect to find in the various regions of our survey.
We assume a thin, exponential disk, where the stellar density follows a 
radial exponential law:

\begin{equation}
N_{*}=N_{0}e^{-(R-R_{0})/\alpha_{R}},
\end{equation}

and an exponential dust distribution:

\begin{equation}
a_{K}(R)=a_{K,0}e^{-(R-R_{0})/\alpha_{R}},
\end{equation}

where R is the Galactocentric distance, R$_{0}$ is the solar Galactocentric
distance of $8.0$~kpc, $\alpha_{R} = 3.0$~kpc from Kent et al. (1991).
We take a$_{K,0}$ to be $0.00008$~mag~pc$^{-1}$ for consistency with 
extinction measurements near the Galactic Center (\cite{CWG90}).
This model predicts a maximum extinction at the Galactic Center of 
$A_{K} \sim 3.4$.

The WR progenitors, the O stars, should follow this exponential law, but we 
must also consider the metallicity dependence for WR formation 
(\cite{MM94}, which modifies $\alpha_{R}$.

\begin{equation}
WR/O = 0.13 Z/Z_{0}
\end{equation}

For young stars and emission nebulae, the metallicity variation within 
the Galactic disk follows an exponential drop-off. Smartt \& Rolleston 
(1997) found an oxygen gradient in the Galactic disk of:

\begin{equation}
12 + log (O/H) = -0.07 \times R_{g}(kpc) + 9.4(dex)
\end{equation}

This leads us to the final equation for WR stellar density as a function
of longitude and distance from the galactic center:

\begin{equation}
N_{*}=N_{0,WR}e^{-(R-R_{0})/\alpha_{WR}},
\end{equation}

where $\alpha_{WR}$ has the value 2022 pc and N$_{0,WR}=2.2$~kpc$^{-2}$
from studies in the local solar neighbourhood (\cite{AM91}, \cite{CV90}),
\cite{M96}). This relation predicts a total of $\sim 2900$ WR stars 
in the Galaxy (compared with $\sim 2500$ in Shara et al. 1999).

We now have an expression for WR density and dust extinction, the next
ingredient is absolute $K$ magnitudes for WR stars, which are not well--known.
From previous spectral analyses, WC absolute $K$ magnitudes range from 
-3.8 to -5.7 (\cite{Setal01}, \cite{Detal00}, \cite{HM99}, \cite{Cetal02}). 
For late-type WCs, which are predominant at high metallicity, absolute
K magnitudes range from -3.8 to -5.0. The absolute $K$ magnitudes of WR134 
and WR136, both of WN4$-$5s type, are -5.47 and -5.64, respectively 
(\cite{CS96}). The Galactic star WR105 (WN9h) has an absolute magnitude of 
-6.0 derived from its apparent $K$ magnitude, its distance and extinction
at $K$ (\cite{Cetal92}).

Given that WN and WC stars come from a range of initial masses, they
should also have a range in absolute magnitude. Additionally, the 
luminosity of the WNL phase is greater than that of the WC phase
while at comparable temperatures. As expected, it appears from 
observations that WN and WC stars have a range in absolute magnitude, and 
the WN star population extends to brighter magnitudes.
Let us assume a number distribution according to a Salpeter law, 
that WN stars are distributed between K~$-4.0$ and $-7.0$, WC stars
between $-3.0$ and $-6.0$, and that the WC/WN ratio is 1.2 in the inner 
galaxy (from the empirical relation for WC/WN and O/H in the Local Group,
\cite{MJ98}).

\subsection{Smooth Model Prediction}

\begin{figure}
 \centering
 \includegraphics[width=8.5cm]{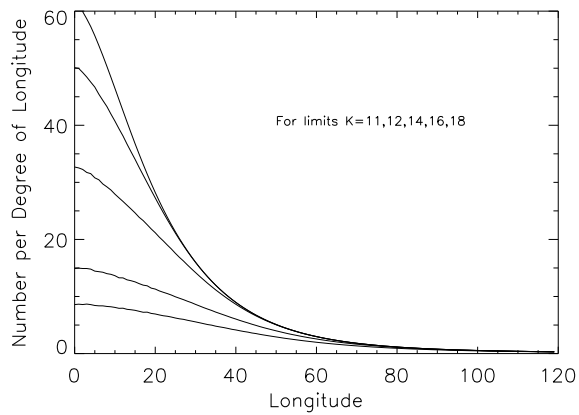}
 \caption{The number of WR stars expected from our simple model
per degree of longitude as a function of Galactic Longitude.}
 \label{longwr}
\end{figure}

With this model, we can make an estimate of the number of 
candidates we can expect in our data set. In Figure \ref{longwr} we show 
the number of candidates
expected per degree of Galactic longitude as a function of longitude,
for varying apparent magnitude limits. 

Our images cover $\sim 3$ degrees of longitude, but $\sim 15\%$ of that is
lost by being on the edges of our strips, and 3 strips are of very low
signal-to-noise due to clouds. This gives us a total of 2.3 degrees
of longitude, covered between $\pm 0.5$ galactic latitude.

Within that region our model predicts 18 WR stars brighter than K~$=11$, 
and 34 brighter than
K~$=12$, or 16 additional stars. Our completeness tests indicate that we 
detect $95\%$ of emission line stars brighter than K$=11$, and $75\%$ 
brighter than K$=12$. According to this, we should detect a total of 28 
WR stars in this region. 
In our second set of data, taken in 2000 at $l=316$, we cover about one
degree of longitude between $b \pm 1.0$, and expect 4 WR stars.

So far, we have observed all our high priority candidates in the GC
region. We have not yet observed any candidates in the
Centaurus--Scutum region. In the GC region, we have found four new WR
stars. These four stars should be added to those already found in the
region through pointed observations. The majority are found in two
young clusters, the Arches cluster (Nagata et al. 1995; Cotera et al. 1996),
and the Quintuplet cluster (Figer et al. 1999), though some stars have been
found outside clusters as in the present case (Cotera et al. 1999). A
full discussion of our detection efficieny and description of the
newly discovered WR stars will be given in our upcoming paper on
spectroscopic follow--up observations. 

Preliminary spectroscopic results are given by \cite{Hetal02}.

\section{Other Emission--line Objects}

Along with WR stars, we may detect other emission line objects in our 
survey. As mentioned previously, LBVs are related to WRs, and also have
strong emission lines in the infrared. Be and B[e] stars have Br$\gamma$
lines, but much weaker than for WRs or LBVs. Therefore
only the strongest emitters may be detected. 

Another possibility is WR central stars of planetary nebulae. The stars
themselves have very weak lines in this region, but their surrounding
nebulae have lines of He I at 2.06 $\mu$m and B$\gamma$ at 2.166 $\mu$m.

We can also detect very young O and B stars still enshrouded in their
natal gas and dust. These objects form compact, or ultra--compact H~II
regions, and later, when they are more revealed, can exhibit emission
from circumstellar disks. They will be visible in nebular emission
lines of Br~$\gamma$ and He~$\ion I 2.06 \mu$m (\cite{BD99b},
\cite{HLR02}). Along with the WR stars which we have already detected
(see above), we have identified several compact Br$\gamma$
sources. These will be discussed in our following paper detailing the
spectroscopic observations of our candidate objects.

\section{Summary}

We have completed a near--infrared survey of approximately three
square degrees toward the inner Galaxy including a larger region
centered on the Galactic center (GC) and a smaller region towards the
Centaurus--Scutum arm. Our survey uses four line emission filters
and three continuum filters in the $K-$band. The need for a continuum filter
on both the red and blue side of each line is driven by the large and
variable extinction toward the inner Galaxy.

We have recently completed spectroscopic follow-up for imaging data
taken in 1996--1998 centered on the GC. We discovered four new WR
stars. The complete results, including spectra, coordinates, and
finding charts will be published in an upcoming paper (\cite{Hetal03},
in prep). Preliminary spectroscopic results are given by \cite{Hetal02}.




\acknowledgements{N.H. acknowledges and thanks the ESO Studentship Programme
and the Wisconsin Space Grant Consortium Graduate Fellowship Program.
N.H. would also like to thank the University of Wisconsin Graduate School 
for partial support. P.S.C. appreciates continuous 
support from the NSF. The authors would like to thank
Ted LaRosa and Michael Nord for providing their 90cm radio image, and
P. Eenens for providing his K-band WR spectra. We would like to acknowledge
the continuing excellent support of the CTIO mountain staff.
This research has made use of the NASA/IPAC Infrared 
Science Archive, which is operated by the Jet Propulsion
Laboratory, California Institute of Technology, under contract with the 
National Aeronautics and Space Administration.}


\begin{thebibliography}{}

\bibitem[Armandroff \& Massey 1991]{AM91} Armandroff, T. E., \& Massey, P. 1991, AJ, 102, 927

\bibitem[Blum et al. 1995]{BDS95} Blum, R. D., DePoy, D. L., \& Sellgren, K. 1995, ApJ, 441, 603

\bibitem[Blum \& Damineli 1999b]{BD99b} Blum, R. D., \& Damineli, A. 1999b, ApJ, 512, 237

\bibitem[Blum \& Damineli 1999a]{BD99a} Blum, R. D., \& Damineli, A. 1999a, in: K. A. van der Hucht, G. Koenigsberger, \& P. R. J. Eenens (eds.), Wolf-Rayet Phenomena in Massive Stars and Starburst Galaxies, Proc. IAU 193 (San Francisco: ASP), 472

\bibitem[Blum et al. 2001]{Betal01} Blum, R. D., Schaerer, D., Pasquali, A., Heydari-Malayeri, M., Conti, P. S., \& Schmutz, W. 2001, AJ, 122, 1875

\bibitem[Catchpole et al. 1990]{CWG90} Catchpole, R. M., Whitelock, P. A., \& Glass, I. S. 1990, MNRAS, 247, 479

\bibitem[Churchwell et al. 1992]{Cetal92} Churchwell, E., Bieging, J. H., van der Hucht, K. A., Williams, P. M., Spoelstra, T. A. Th., \& Abbott, D. C. 1992, ApJ, 393, 329

\bibitem[Conti 1976]{C76} Conti, P. S. 1976,in: Ministere de l'Education National, Fonds National de la Recherche Scientifique and Universite de Liege, Colloque International d'Astrophysique 20th, vol. 9, 193

\bibitem[Conti 1991]{C91} Conti, P. S. 1991, ApJ, 377, 115

\bibitem[Conti \& Vacca 1990]{CV90} Conti, P. S., \& Vacca, W. D. 1990, AJ, 100, 431

\bibitem[Cotera et al. 1996]{Cetal96} Cotera, A. S., Erickson, E. F., Simpson, J. P., Colgan, S. W. J., Allen, D. A., \& Burton, M. G. 1996, ApJ, 461, 750

\bibitem[Cotera et al. 1999]{Cetal99} Cotera, A. S., Simpson, J. P., Erickson, E. F., Colgan, S. W. J., Burton, M. G., \& Allen, D. A. 1999, ApJ, 510, 747 

\bibitem[Crowther \& Smith 1996]{CS96} Crowther, P. A., \& Smith, L. J. 1996, A\&A, 305 541

\bibitem[Crowther et al. 2002]{Cetal02} Crowther, P. A., Dessart, L., Hillier, J., Abbott, J., \& Fullerton, A. 2002, A\&A, 392, 653

\bibitem[Damineli et al. 1997]{Detal97} Damineli, A., Jablonski, F., de Freitas, L. C., \& de Freitas-Pacheco, J. A. 1997, PASP, 109, 633

\bibitem[Dessart et al. 2000]{Detal00} Dessart, L., Crowther, P. A., Hillier, J. D., Willis, A. J., Morris, P. W., \& van der Hucht, K. A. 2000, MNRAS, 315, 407

\bibitem[Figer et al. 1997]{FMM97} Figer, D. F., McLean, I. S., \& Morris, M. 1997, ApJ, 486, 420

\bibitem[Figer et al. 1999a]{Fetal99a} Figer, D. F., McLean, I. S., \& Morris, M. 1999a, ApJ, 514, 202

\bibitem[Figer et al. 1999b]{Fetal99b} Figer, D. F., Kim, S. S., Morris, M., Serabyn, E., Rich, R. M., \& McLean, I. S. 1999b, ApJ, 525, 750

\bibitem[Hanson et al. 2002]{HLR02} Hanson, M. M., Luhman, K. L., \& Rieke, G. H. 2002, ApJSS, 138, 35

\bibitem[Heckman et al. 1998]{Hetal98} Heckman, T., Robert, C., Leitherer, C., Garnett, D., \& van der Rydt, F. 1998, ApJ, 503, 646

\bibitem[Heckman et al. 2001]{Hetal01} Heckman, T., Sembach, K., Meurer, G., Leitherer, C., Calzetti, D., \& Martin, C. 2001, ApJ, 558, 56

\bibitem[Hillier \& Miller 1999]{HM99} Hillier, J. D., \& Miller, D. L. 1999, ApJ, 519, 354

\bibitem[Homeier et al. 2002]{Hetal02} Homeier, N. L., Blum, R. D., Conti, P. S., Pasquali, A., \& Damineli, A. 2002, in K.A. van der Hucht, A. Herrero \& C. 
Esteban (eds.), A Massive Star Odyssey, from Main Sequence to Supernova,
Proc. IAU Symp. No. 212 (San Francisco: ASP), in press

\bibitem[Homeier et al. 2003]{Hetal03} Homeier, N. L., Blum, R. D., Pasquali, A., Conti, P. S., \& Damineli, A. 2003, in prep

\bibitem[Kent et al. 1991]{KDF91} Kent, S. M., Dame, T. M., \& Fazio, G. 1991, ApJ, 378, 131

\bibitem[Krabbe et al. 1995]{Ketal95} Krabbe, A., Genzel, R., Eckart, A. et al. 1995, ApJ, 447, 95

\bibitem[Langer et al. 1999]{Letal99} Langer, N., Garc\'{i}a-Segura, G., Mac Low, M.-M. 1999, ApJ, 520, 49L

\bibitem[LaRosa et al. 2000]{Larosaetal00} LaRosa, T. N., Kassim, N. E., Lazio, T., Joseph, W., \& Hyman, S. D. 2000, AJ, 119, 207

\bibitem[Leitherer, Robert, \& Drissen 1992]{LRD92} Leitherer, C., Robert, C., \& Drissen, L. 1992, ApJ, 401

\bibitem[Maeder \& Conti 1994]{MC94} Maeder, A., \& Conti, P. S. 1994, ARA\&A, 32, 227

\bibitem[Maeder \& Meynet 1994]{MM94} Maeder, A., \& Meynet, G. 1994, A\&A, 287, 803

\bibitem[Maeder \& Meynet 2000]{MM00a} Maeder, A., \& Meynet, G. 2000, ARAA, 38, 143

\bibitem[Maeder \& Meynet 2001]{MM01} Maeder, A., \& Meynet, G. 2001, A\&A, 373, 555

\bibitem[Martin 1999]{M99} Martin, C. L. 1999, ApJ, 513, 156

\bibitem[Massey 1996]{M96} Massey, P. M. 1996, in Wolf-Rayet Star in the Framework of Stellar Evolution, ed. J.-M. Vreux, A. Detal, D. Fraipont-Caro, E. Gosset, \& G. Rauw (Li\`{e}ge:Inst. d'Astrophys., Univ. Li\`{e}ge), 361

\bibitem[Massey \& Johnson 1998]{MJ98} Massey, P., \& Johnson, O. 1998, ApJ, 505, 793

\bibitem[Meynet \& Maeder 2000]{MM00b} Meynet, G., \& Maeder, A. 2000, A\&A, 361, 101

\bibitem[Nagata et al. 1995]{Netal95} Nagata, T., Woodward, C. E., Shure, M. \& Kobayashi, N. 1995, AJ, 109, 1676

\bibitem[Oey \& Clarke 1997]{OC97} Oey, M. S., \& Clarke, C. 1997, MNRAS, 289, 570

\bibitem[Oey et al. 2001]{Oetal01} Oey, M. S., Clarke, C., \& Massey, P. 2001,in: Klaas S. De Boer, Ralf-Juergen Dettmar, \& Uli Klein (eds.), Dwarf Galaxies and their Environments, 40th meeting of the Graduiertenkolleg ``The Magellanic Clouds and other dwarf galaxies'' (Aachen: Shaker Verlag), 181

\bibitem[Pasquali et al. 1997]{Petal97} Pasquali, A., Langer, N., Schmutz, W., Leitherer, C., Nota, A., Hubeny, I., \& Antony, F. J. 1997, ApJ, 478, 338

\bibitem[Schechter et al. 1993]{SMS93} Schechter, P., Mateo, M., \& Saha, A. 1993, PASP, 105, 1342

\bibitem[Shara et al. 1999]{Setal99} Shara, M. M., Moffat, A. F. J., Smith, L. F., Niemela, V. S., Potter, M., \& Lamontagne, R. 1999, AJ, 118, 390

\bibitem[Smartt et al. 2001]{Setal01} Smartt, S. J., Crowther, P. A., Dufton, P. L., Lennon, D. J., Kudritzki, R. P., Herrero, A., McCarthy, J. K., \& Bresolin, F. 2001, MNRAS, 325, 257S

\bibitem[Smartt \& Rolleston 1997]{SR97} Smartt, S. J., Rolleston, W. R. J. 1997, ApJ, 481L, 47

\bibitem[Smith et al. 1998]{Setal98} Smith, L. J., Nota, A., Pasquali, A., Leitherer, C., Clampin, M., Crowther, P. A. 1998, ApJ, 503, 278


\end{thebibliography}
\end{document}